# Integrated Optofluidic Sensor for Coagulation Risk Monitoring in COVID-19 Patients at Point-of-Care

*Robin Singh,[1,2], Alex Benjamin[1], Galit Frydman, Lionel Kimerling[3,4,5], Anu Agarwal[3,4,5] and Brian W Anthony[1,2]*

*[1]Department of Mechanical Engineering, [2]MIT nano,*

*[3]Department of Materials Science and Engineering,*

*[4]Microphotonics Center, [5]Materials Research Laboratory, Massachusetts Institute of Technology, Cambridge, MA, 02139, USA*

## Abstract:

While the pathophysiology underlying the COVID-19 infection remains incompletely understood, there is growing evidence to indicate that it is closely correlated to hypercoagulation among severely ill patients. Doctors may choose for use anti-coagulation doses to treat the patients at intensive care units. A rapid, easy, and low-cost solution to monitor the coagulation status at the point-of-care may help with treatment by enabling the administration of controlled doses of medication to patients and to understand the disease's underlying pathophysiology. Thromboelastography, the clinical standard is accurate; it suffers from limited portability and low sensitivity when miniaturized to handheld form factor. In the article, we summarize research helping to advance towards an integrated optofluidic device combining microfluidics and photonic sensor technology. Microfluidics are used to perform blood pre-processing, and a photonic sensor measures blood coagulation status in real time readout on the device itself. These techniques make it portable and scalable, potentially serving as technology foundation for the development of a disposable sensor for point-of-care diagnostics in COVID-19 patients and coagulopathy in general.

# 1. Introduction

The rapid spread of COVID-19 has resulted in a global pandemic with very high rates of morbidity and mortality. As of May 22, 2020, it has infected more than 5 million people in different parts of the world. Various disease models have estimated the overall mortality rate to be between 4.3% to 14.6 %. A leading contributor to this mortality rate is the development of progressive bilateral pneumonia which leads to acute respiratory distress syndrome (ARDS) and eventual death. It is believed that the surface proteins of the COVID-19 virus bind to ACE2 receptors on the host cell and subsequently invade it via clathrin-mediated endocytosis [1,2]. ACE2 receptors are broadly expressed in the vascular endothelium, respiratory endothelium, alveolar monocytes, and macrophages. As a result, the virus is able to effectively replicate in the upper and lower respiratory tracts, eventually attacking target organs that express ACE2 (heart, kidney, gastrointestinal tract). During disease progression, the exaggeration results not only from the direct viral infection but also from the immune-mediated injury caused by COVID-19. Multiple studies have reported two distinctive features among patients: progressive inflammation and hypercoagulation [3-8].

In a recent study, clinical researchers reported that elevated levels of fibrin degradation D-dimers were indicative of poor prognosis of the disease [9]. Working under the hypothesis that coagulopathy is closely correlated to the pathogenesis of COVID-19, multiple post-mortem studies have reported the presence of drastic microvascular changes in the lungs, including disseminated micro-thrombi and hemorrhagic necrosis [10]. Other studies suggest that COVID-19 increases the risk of deep vein thrombosis and pulmonary embolisms. Recently, Maxime et al. reported the occurrence of venous thromboembolism (VTE) among admitted COVID-19 patients [11].

The coagulation response of the human vasculature system against external stimulus or injury is well known; it is unclear how COVID-19 triggers this process [3]. One hypothesis is that COVID-19 prompts the release of chemical signals from the immune cells that ramp up inflammation and blood coagulation [3,6,7]. Consequently, doctors are relying on anti-coagulative medication to increase the survival rates in admitted cases. Paranjpe et al. published their findings on the association of treatment with anti-coagulation dose with in-hospital survival of the patients [8]. The effect of anticoagulation treatment is more pronounced among ventilated patients: 62.7 percent of intubated patients who were not treated with anticoagulants died as compared to 29.1 percent for intubated patients who were treated with anticoagulants [7].

Given the critical abnormality of coagulopathy among COVID-19 patients, we believe a rapid, quantitative, portable and low-cost device for hemostasis monitoring at the point-of-care is needed. This would enable precise and personalized anticoagulation response monitoring among critically ill patients [12]. A bed-side test should be minimally invasive, and require a low blood volume per test. This is especially true for patients who are assisted by extracorporeal devices such as hemodialysis machines, membrane oxygenation systems, and mechanical circulatory system support systems [12,14]. Those patients require monitored anticoagulation to prevent blood clotting in the transport lines and pumps of these devices, ex vivo [14-16].

Many tests and devices have been developed to assess and monitor blood coagulation and platelet functions [15-25]. The current standard is based on thromboelastography (TEG); TEG provides a rapid assessment of hemostasis in the blood by measuring shear elasticity and the dynamics of clot formation [25-29]. While it provides information on platelet function, coagulation formation, and fibrinolysis, it generally requires expensive and large equipment to perform the tests. The method is not currently portable and cannot be used in most point-of-care settings; this is uniquely true in resource constrained or high throughput settings (e.g. urgent care clinics) [28,30]. Other tests and devices have also been developed to assess blood clotting in vitro, for instance, assays for bleeding time, activated clotting time, and platelet aggregometry [19-21]. But, they are unable to provide accurate predictions of thrombotic and bleeding risks. Often, they are inaccurate and have poor sensitivity in detecting the different stages of blood coagulation cascade [ 23-25].

In this manuscript, we describe a simple optofluidic approach for real-time monitoring of coagulation status among severely-ill COVID patients. A device based on this approach would permit rapid determination of the effectiveness of anti-coagulant treatment and allow medical professionals to tailor dosage to meet patient-specific response. The proposed approach is minimally invasive and allows for the development of scalable diagnostic and treatment solutions in the current pandemic.  The presented optofluidic sensor module system includes a microfluidic device that is integrated into a photonic sensor [31-33]. The microfluidic chip performs passive separation of blood plasma, which then interacts with the sensor module in the presence of the coagulation/anti-coagulation dose to monitor the clotting of blood. The sensor module is composed of SiN based photonic micro-ring resonators to detect [34] and quantify the different stages of the blood coagulation cascade. The proposed approach could be made into a low cost, rapid, and accurate device solution.

## 2. Results

In general, blood is a multiphase fluid made up of different cells in plasma [35,36]. It consists of erythrocytes (red blood cells, RBCs), leukocytes (white blood cells, WBCs), and thrombocytes. By volume (v/v %), the composition is about 45 to 50% of red blood cells, 1% of white blood cells, <1% of platelets, and 55% of plasma fluid [37]. The process of blood coagulation is a complex cascade that involves platelet cells, plasma protein fibrinogen, and other clotting factors that form fibrin strands to strengthen the platelet plug [38]. Larger cells such as RBCs and WBCs are absent in this process. Moreover, they are significantly larger (about 8 to 12 μm) than platelet cells, which are involved in the coagulation cascade and can interfere with the coagulometric detection method proposed here. It is necessary to separate the plasma before detecting the blood coagulation cascade [39-41].

Conventionally, plasma separation is performed using a benchtop centrifuge; this is infeasible for integration into a point of care biosensor [35]. With microfluidics, however, there are approaches to miniaturize such traditional separation methods [42,43]. Various methods have been proposed to separate the plasma from blood. Broadly, they can be classified as passive separation and active separation methods [35, 42-45]. To reduce the complexity, we choose a passive approach for separating the platelet-rich plasma from blood.

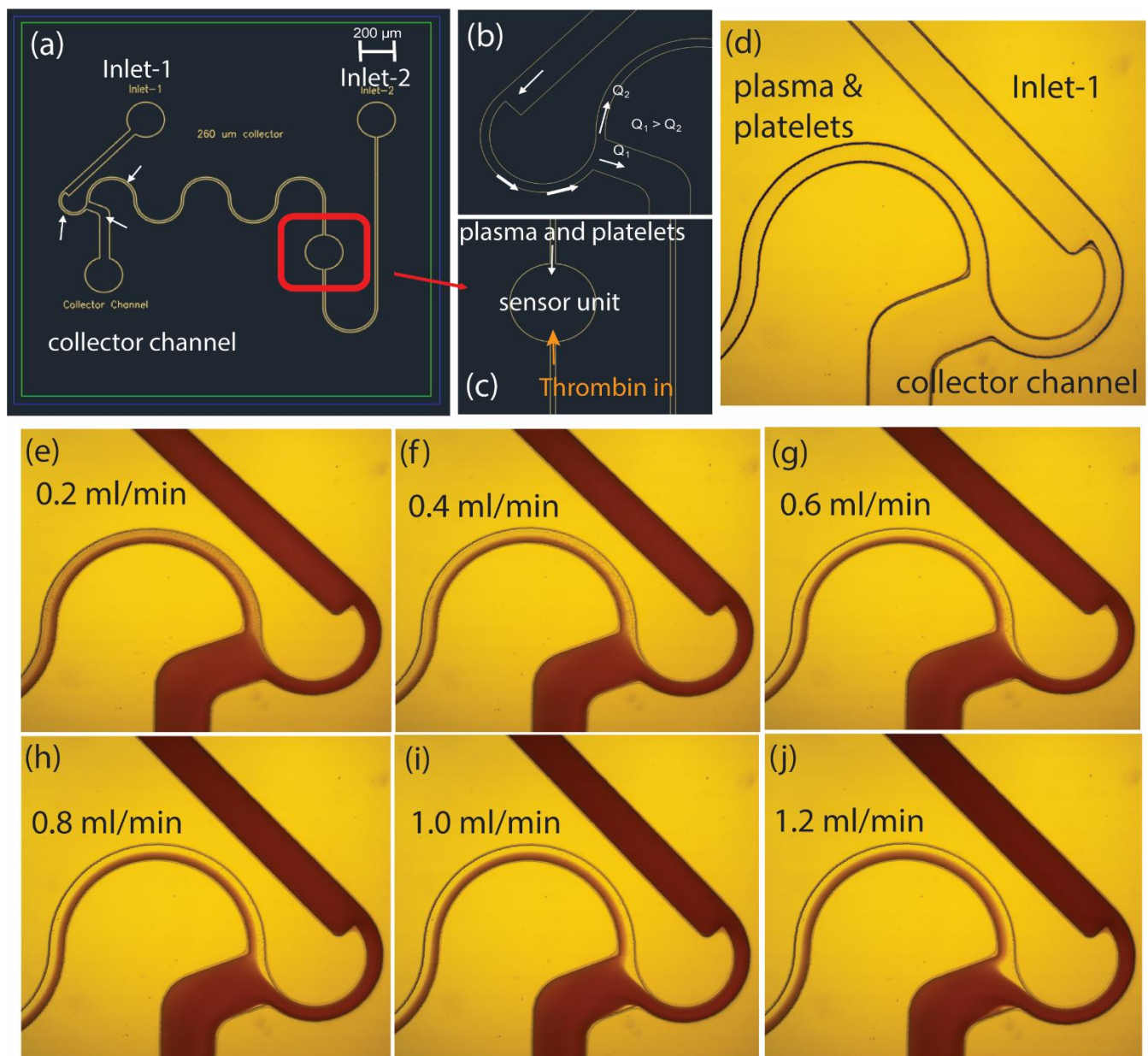

*Figure-1: Design of a microfluidic device for on-chip plasma separation. (a) CAD design of the microfluidic device. The whole blood sample is input through inlet-1 with a given flow rate. Collector channel collects the RBC rich blood component. (b) The blood is allowed to flow through a curvature that later ends with a bifurcation. The bifurcating daughter channels are designed in such a way that there is flow rate different in the channels. Due to the Zweifach Fung bifurcation effect combined with the centrifugal effect, RBCs prefer flowing in the channel with a higher flow rate. (c) Plasma component of the blood passes on to the serpentine like channel. Coagulating agent flows in through the inlet-2 and mixes with the plasma at the sensor unit that monitors the coagulation in real time. (d-j) Comparison of the microfluidic device performance for different flow rate of the whole blood. (d) Picture of the empty microfluidic device. (e-j) Plasma separation from the whole blood for flow rate of 0.2 ml/min, 0.4 ml/min, 0.6 ml/min, 0.8 ml/min, 1 ml/min, 1.2 ml/min respectively.*

We combine the prominent effects of Fahraeus, Zweifach Fung Bifurcation and inertia to separate larger cells from whole blood under test [42, 45-47]. We fabricate a curved microfluidic channel with a bifurcation which allows fluids to flow at different rates in the daughter channels. Fig. 1-a shows the schematic of the design. Curvature in the channel results in a centrifugal force acting on the inertia-dominated red blood cells (RBCs). The forces result in the accumulation of RBCs towards the periphery of the channel. The channel ends with a bifurcation into two channels. The dimensions of these two channels are selected to create different flow rates. RBCs flowing through the channel accumulate towards the periphery. Experiencing the Zweifach Fung effect at the bifurcation site, RBCs preferentially flow towards the larger collector channel [35]. This leaves the plasma rich component in the serpentine structure, where it combines with the coagulating agent at the sensor site. An optical sensor performs measurements on the sample and monitors the coagulation process in real-time.

Blood is a dielectric medium which responds to an external electromagnetic field by changing the orientation of its dipoles. The polarization in the blood sample can be orientational or deformational in nature [48-50]. Polarization changes in the optical medium are represented in terms of its refractive index ($\eta$). We define $\eta$ as the square root of the relative dielectric of the medium . During the coagulation cascade, due to cellular activation and molecular interactions, the dielectric constant and therefore refractive index of the blood changes. If we model the change in the refractive index (or dielectric constant) as [50-52]

$$n_{plasma} = \frac{n_{plasma}^{\text{int}}}{1 \mp e^{\pm k(t-t_0)}}$$

(1)

where, $n_{plasma}^{\text{int}}$ is the refractive index of the blood in its normal states (in our case, TS-0 when the coagulation or anticoagulation agent is introduced) .The progressive change in the refractive index of blood is manifested as changes in resonant peaks ($\lambda$) of the optical ring resonators. The resonant wavelength for mode, $m$ of the photonic ring resonator with circumferential length, l is given by [32-33, 53]

$$\lambda_{res} = \frac{n_{eff} L}{m}$$

(2)

The relation between change in $n_{eff}$ and $\lambda_{res}$ is established by the first order perturbation theory as

$$\Delta\lambda_{res} = \frac{\left[\left(\frac{\partial n_{eff}}{\partial n_{plasma}}\right)_{\lambda_{res}, n_{plasma}^0} \Delta n_{plasma} + \left(\frac{\partial n_{eff}}{\partial \lambda}\right)_{\lambda_{res}, n_{plasma}^0} \Delta\lambda_{res}\right] L}{m}$$

(3)

where $\Delta\lambda_{res}$ is the change in resonance wavelength, $n_{blood}$ is the effective index of the blood surrounding the photonic waveguides, $n_{eff}$ is the effective index of the photonic waveguide, and m is the resonant mode. If we define the group effective index, $n_g$ as below,

$$n_g = n_{eff} - \lambda_0 \frac{dn_{eff}}{d\lambda}$$

(4)

Equation -3 can be expressed as,

$$\Delta\lambda_{res} = \frac{\Delta n_{plasma} \lambda_{eff}}{n_g}$$

(5)

A change in effective index of blood can be calculated by perturbation theory using the equation,

$$\Delta_{plasma} n_{eff} = c \int \Delta \varepsilon_{plasma} E_v E_v^* dx dy \qquad (6)$$

We capture the change in the effective index of the blood via change in the resonance peak of the photonic ring resonator.

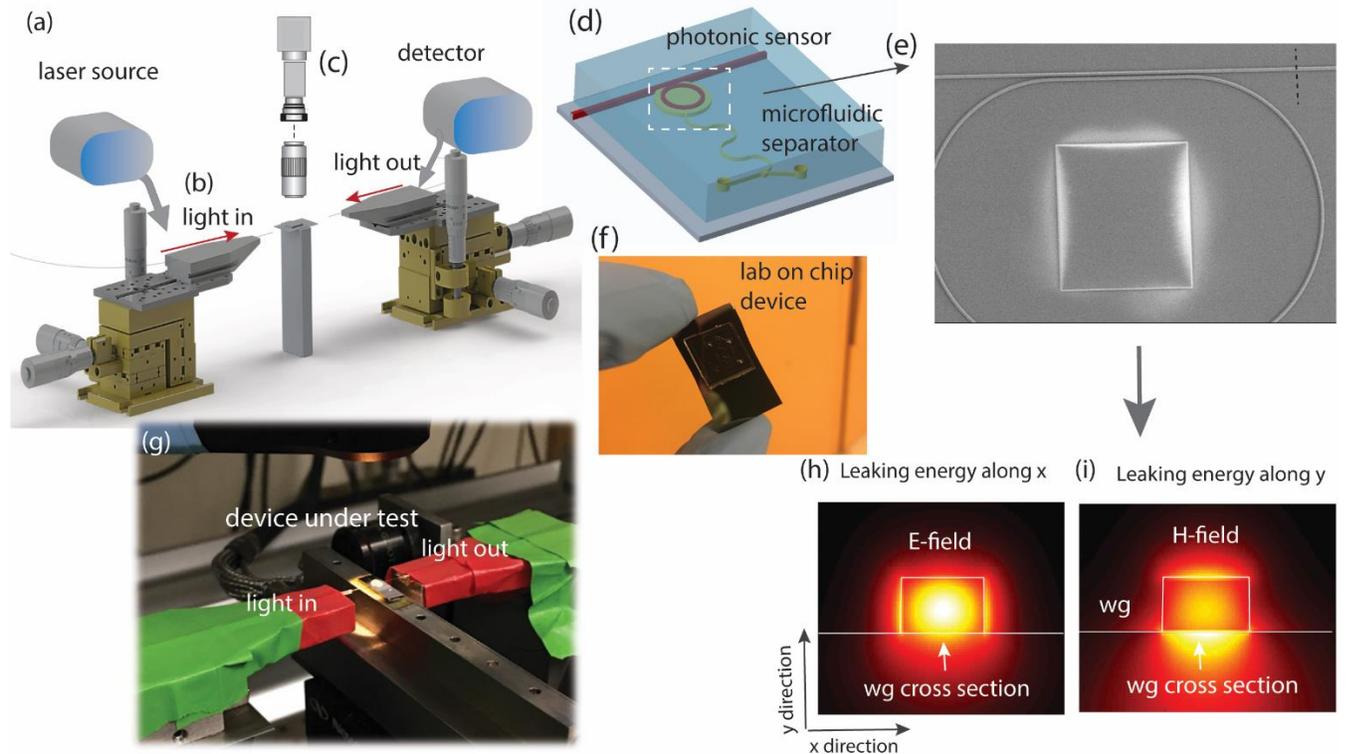

*Figure-2: (a) Camera picture of Lab-on-chip optofluidic device. (b) Coupling of light through lensed fiber tips resting on the 3-axis stages with the photonic sensor module. The alignment of the fiber tips and microfluidic device is visualized through an inverted zoom lens setup. (c) Coupling of microfluidic channels to the sensor module. (d) SEM image of the fabricated photonic chip. (e,f) Evanescent leakage of light energy along x-direction and y-direction. The leaking energy interacts the clotting blood flowing in the optofluidic channel. (i) Time step (TS-0) when the coagulation factor, Thrombin is not added to the blood plasma rich in platelet count. (j-l) When the thrombin is added; plasma starts to form cross-linked fibrin strands measured after (j) Time step (TS-1) after 40 seconds (k) Time step (TS-2) after 120 seconds and (l) Time step (TS-3) after 160 seconds.*

## 3. Discussions

We first evaluate the efficacy of RBC separation from the whole blood sample. To determine the optimal flow rate for the blood in our device, we experiment with different flow rates. The blood is purchased from Research Blood Components, LLC, Boston, MA. Fig. 1 shows the device and representative images from the experiments. The hematocrit concentration of blood is 38 %. It refers to the concentration of red blood cells in the whole blood in v/v %. Fig. 1e-1j compare the device's plasma separation ability for different flow rates. From Fig. 1 e, it is evident that the 0.2 ml/min is not sufficient to provide the required centrifugal force and flow rate difference to separate plasma from the blood, and we can see a significant red tinge (RBCs) in the channel going towards the sensor. As we increase the flow rate to 0.4 ml/min and beyond, the separation efficiency increases. We can see the improved performance from the clear plasma in the sensor channel. However, there is no significant change in separation efficiency as we go beyond 0.6 ml/min. As the flow rate increases further, a constriction in the flow is observed (Fig. 1j), thereby reducing the separation yield. Therefore, we select the flow rate for our sensor to 0.6 ml/min.

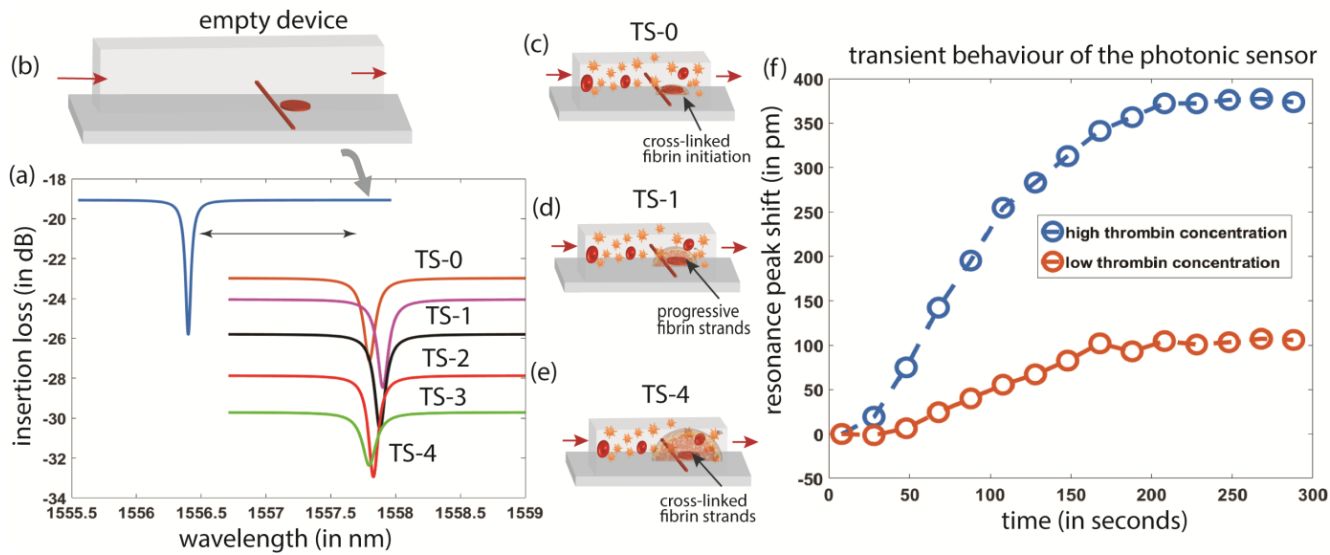

*Figure 3: Figure representing the response of ring resonator in response to addition of Thrombin and coagulation. (a) Shift in resonance peak at different time stamp (TSs). (b,c,d,e) Definition of different time stamps. (f) Transient variation of ring resonators when different concentration of Thrombin was added to the device. Higher concentration was roughly twice of that of lower concentration.*

After optimizing the flow rate of blood in the microfluidics, we perform experiments to characterize the response of on-chip resonator cavities to coagulating blood. We use thrombin as a coagulating agent added to the blood plasma via inlet-2 (shown in Fig. 1a). Fig. 2 shows the experimental setup used to couple in light to the optofluidic chip through the lensed fiber tips. The inverted microscope setup assists the alignment with a zoom lens. In general, the propagation of light energy in the photonic waveguide is accompanied by co-propagation in the surrounding medium, labeled as evanescence, which interacts with the clotting blood to change the effective index of the photonic waveguide. As described in equation-3, it results in a change in the resonant peak of the ring resonators. Fig. 3-a shows the sensor response to the coagulation at different timestamps (TS-1, 2, 3 and 4). We observe a significant redshift (shift towards higher wavelength) in the sensor as thrombin is added. This is attributed to the fact that the addition of thrombin increases the effective index of refraction significantly. Further, as the blood starts clotting, we observe a blue shift (shift towards lower wavelengths) in the resonance peak. The reversal in direction of the shift could be due to the change effective index of the medium as the cross link fibrin strands continue to grow in the clotting blood (as shown in Fig. 3- c-e). The measurements are taken for about 3 minutes and repeated for two independent devices, all yielding similar results. Fig. 3f shows the transient response of the sensors as the blood plasma coagulates over time. We test it on two different devices with different Thrombin concentrations. This suggests that the approach can be used to detect dynamic progression of blood coagulation. More controlled experiments remain, to map different stages of blood clotting with the progressive shift in resonance peak and how it effects its effective index.

## 4. Conclusion

In this manuscript, we present an opto-fluidic approach for coagulation risk monitoring in COVID-19 patients at point-of-care. With the growing pandemic of COVID-19, causing thousands of deaths worldwide, it is customary to scale up the intensive care unit for efficient monitoring of in-hospital patients. There is mounting evidence to show that COVID-19 viral infection leads to hypercoagulation, and thus there is a need to perform qualitative and quantitative measurements of coagulopathy among severely ill patients. A proposed device based on the proposed approach may offers a low cost, rapid and easy method to monitor coagulation using a small test volume. The device combines microfluidic and photonic sensing technology. Microfluidics performs passive separation of the platelet-rich plasma from the whole blood; the photonic sensor detects the change in states of the clotting blood. The approach can provide real-time readouts for representing the alterations in coagulation status ex-vivo, and is more reliable than standard clotting assays. The existing clotting assays may distort the blood sample and can significantly affect the coagulation process. The presented method allows us to scale a disposable technology for personalized diagnostics and treatment, providing real-time observation of anticoagulation therapy for COVID-19 patients. Future work remains in performing more controlled experiments to map different coagulation stages with the progressive resonant peak shifts in the photonic sensor.

## 5. Methods

### 5.1 Fabrication

**Microfluidic Chip Fabrication**

Our microfluidic devices are designed on AutoCAD 2018 software (AUTODESK Incorporated., California, U.S.) and we use an SU-8 2075 master template fabricated using standard lithography procedure. The wafers are first cleaned using a piranha cleaning procedure. We bake the wafers at 200-degree C for at least 5 minutes on a metal-top hot plate. Once, the wafer has roughly cooled down to room temperature, SU-8 photoresist is spun coat on it at 2000 rpm. The wafer is baked at 65-degree C for 2 minutes and then 95-degree C for 5 minutes. Now the next step is the exposure. We use a transparency mask from Advance Reproductions Corporation, MA, USA to pattern the SU-8. The transparency mask is printed on the high-resolution printer (5000 dpi gives about 25-micron features). KS Aligner machine is used to expose the wafer with resist for about 2 minutes. Once the wafer with SU-8 is exposed, we post-bake it in the following sequence. The final step in the master mold fabrication is the development. Under hood, PM-Acetate (SU-8 developer) is poured in the Pyrex container and the wafer is developed in the solution for about 5 minutes. With this, our SU-8 master mold is ready.

After this, we use the master mold to make our PDMS micro fluidic channel. PDMS sylgard (Polydimethylsiloxane, Sylgard 184) is obtained from Dow Corning Incorporated. Pre-polymer components are weighed in 10:1 ratio and mixed. SU-8 master mold wafer is placed in a petri dish and PDMS mixture is gently poured to minimize any bubble formation in the process. Finally, the dish is placed in a convection oven with a temperature set at 65-degree C for about 120 minutes. The PDMS is now carefully peeled from the mold wafer and cut into smaller chips ready to use with our sensors.

As a final step, the PDMS microfluidic and the sensor chips are activated using plasma cleaner for 0.2 minutes and are bonded together with proper alignment.

**Photonic Sensor Fabrication**

We fabricate the device with the process flow as shown in Fig. 3. A low pressure chemical vapor deposition (LPCVD) system is used to deposit 400 nm thick silicon nitride ($Si_3N_4$) layer on 6-inch silicon dioxide wafer (3-micron oxide on Si substrate). These thermal oxide wafers are procured from Wafer Pro LLC, CA. Micro ring resonators and waveguides are designed and patterned on silicon nitride on insulator substrates via photolithography and followed by reactive ion etching to define the geometry of the on-chip sensing components. Fluorine chemistry with a gas mixture of CHF3 and CF4 is used in the dry etching step. Fig. 2-d shows the SEM image of the fabricated resonators with its dimensions of the geometry of the on-chip sensing components. Fig.3 shows the step flow of microfabrication used for the photonic sensor.

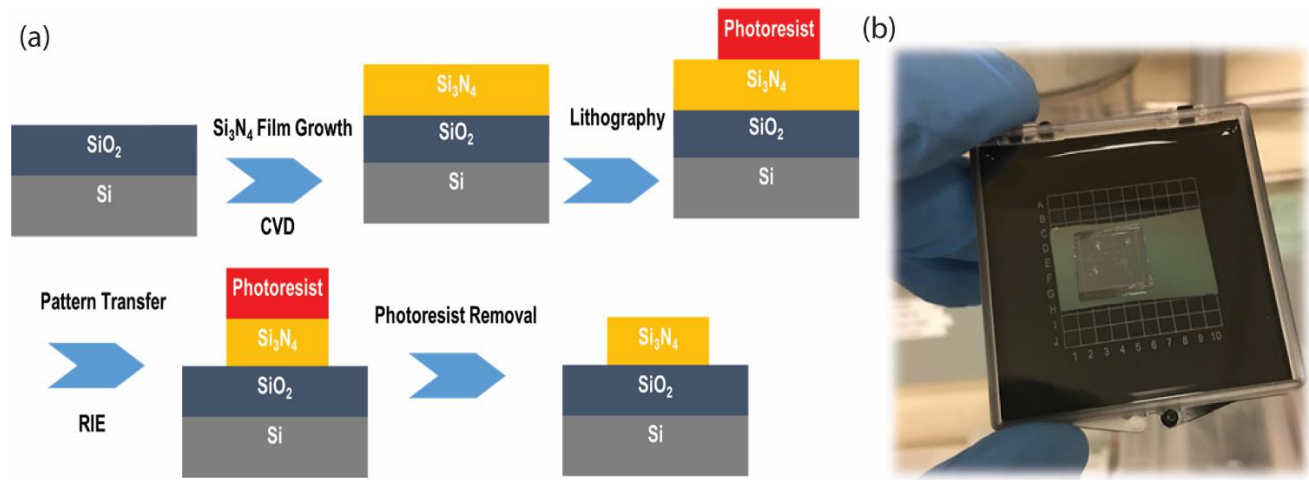

Figure 3: *Schematic showing the process flow to fabricate the Si3N4 based micro race track resonators used in the study. We deposit 400 nm thick $Si_3N_4$ on $SiO_2$+Si wafer. We use photolithography to pattern waveguide structures, followed by pattern transfer using reactive ion etching process. Once the device is patterned, we remove the photoresist by cleaning the wafer with plasma treatment. (b) Camera picture of a fully integrated opto-fluidic device.*

## 6    Acknowledgments

Authors would like to thank Dr. Galit Frydman for her help in carrying out the experiments with the sensor devices.

## 7   Authors Contributions

R. S. performed the experiments, device fabrication and designed the characterization work bench. A.A and B.W.A provided guidance, technical discussions and advised the research. All authors reviewed and edited the manuscript.

## 9    Conflict of Interest

The authors declare that they have no conflict of interest.